\documentstyle[a4,12pt]{article}
\begin{document}

\newcommand{\gsim}{_{\sim}^>}

\newcommand{\lsim}{_{\sim}^{<}}


\noindent
SAGA-HE-87-95

\noindent
June 1995

\bigskip

\bigskip

\noindent
Compressional properties of nuclear matter in the relativistic mean field
theory with the excluded volume effects

\bigskip

\bigskip

\centerline{\bf{H. Kouno*, K. Koide, T. Mitsumori, N. Noda, A. Hasegawa}}

\centerline{Department of Physics, Saga University, Saga 840, Japan}

\centerline{\bf{and}}

\centerline{\bf{ M. Nakano}}

\centerline{University of Occupational and Environmental Health, Kitakyushu
807, Japan}

\bigskip

\centerline{(PACS numbers: 21.65.+f, 21.30.+y )}

\bigskip

\bigskip


\centerline{\bf ABSTRACT}

\bigskip

\noindent
Compressional properties of nuclear matter are studied by using the mean field
theory with the excluded volume effects of the nucleons.
It is found that the excluded volume effects make it possible to fit the
empirical data of the Coulomb coefficient $K_{c}$ of nucleus incompressibility,
even if the volume coefficient $K$ is small($\sim 150$MeV).
However, the symmetry properties favor $K=300\pm 50$MeV as in the cases of the
mean field theory of point-like nucleons.

\bigskip

\bigskip

\noindent
* e-mail address:kounoh@himiko.cc.saga-u.ac.JP

\vfill\eject


One way to determine the incompressibility $K$ of nuclear matter from the giant
monopole resonance (GMR) data is using the leptodermous expansion[1] of nucleus
incompressibility $K(A,Z)$ as follows.
$$  K(A,Z)=K+K_{sf}A^{-1/3}+K_{vs}I^2+K_cZ^2A^{-4/3}+\cdot \cdot
\cdot~~~~~;~~~I=1-2Z/A, \eqno{(1)} $$
where the coefficients $K_{sf}$, $K_{vs}$ and $K_c$ are surface term
coefficient, volume-symmetry coefficient and Coulomb coefficient, respectively.
We have omitted higher terms in eq. (1).
Although there is uncertainty in the determination of these coefficients by
using the present data, Pearson [2] pointed out that there is a strong
correlation among $K$, $K_c$ and the skewness coefficient, i.e., the
third-order derivative of nuclear saturation curve. (See table 1.) Similar
observations are done by Shlomo and Youngblood [3].


\centerline{$\underline{~~~~~~~}$}

\centerline{Table 1}

\centerline{$\underline{~~~~~~~}$}


According to this context, Rudaz et al. [4] studied the relation between
incompressibility and the skewness coefficient by using the generalized version
of the relativistic Hartree approximation [5].
The compressional and the surface properties are studied by Von-Eiff et al.
[6][7][8] in the framework of the mean field approximation of the
$\sigma$-$\omega$-$\rho$ model with the nonlinear $\sigma$ terms.
They found that low incompressibility ($K\approx 200$MeV) and a large effective
nucleon mass $M^*_0$ at the normal density ($0.70\leq M^*_0/M\leq 0.75$) are
favorable for the nuclear surface properties [8].
On the other hand, using the same model, Bodmer and Price [9] found that the
experimental spin-orbit splitting in light nuclei supports $M^*_0\approx
0.60M$.
The result of the generator coordinate calculations for breathing-mode GMR by
Stoitsov, Ring and Sharma [10] suggests $K\approx 300$MeV.

In previous papers[11][12], we have studied the effective nucleon mass $M^*_0$,
incompressibility $K$ and the skewness $K'$ in detail, using the relativistic
mean field theory with the nonlinear $\sigma$ terms [13] and the one with the
nonlinear $\sigma$ and $\omega$ terms [14].
We found that $K=300\pm 50$MeV is favorable to account for $K$, $K_c$, $K_{vs}$
and the symmetry energy $a_4$, simultaneously [11][12].
It was also found that the empirical values of $K$ and $K_c$ in table 1 are not
well reproduced by these model, if $K\lsim 200$MeV [11][12].

On the other hand, the incompressibility $K$, which is calculated in the
framework of the quark-meson coupling (QMC) model is rather small$(\sim
200$MeV) [15][16][17].
The sub-structure or the finite size effect of nucleons may be important in
calculating $K$.
In this paper, we study $M^*_0$, $K$, $K'$, $K_c$ and $K_{vs}$, which can be
calculated in the framework of nuclear matter with aid of the scaling model
[1], by using the relativistic mean field theory with the excluded volume
effects (EVE) of nucleons [18][19][20],  and compare the results with the GMR
data.
We also examine whether the QMC result is reproduced by the EVE model.


We use the relativistic mean field theory with the EVE of nucleons
[18][19][20].
As a Lagrangian, we use the $\sigma$-$\omega$ model with the nonlinear $\sigma$
terms as in ref. [20].
( For a while, we restrict our discussions to the symmetric nuclear matter and
do not consider the $\rho$ meson effects. )
The Lagrangian density consists of three fields, the nucleon $\psi$, the scalar
$\sigma$-meson $\phi$, and the vector $\omega$-meson $V_\mu$, i.e.,
$$ L_{N\sigma\omega} =\bar{\psi}(i\gamma_\mu\partial^\mu
-M)\psi+{1\over{2}}\partial_\mu\phi\partial^\mu\phi -{1\over{4}}F_{\mu\nu}
F^{\mu\nu} +{1\over{2}}m_v^2V_\mu V^\mu  $$
$$ +g_s\bar{\psi}\psi\phi-g_v\bar{\psi}\gamma_\mu \psi V^\mu-U(\phi
)~~~;~~~~~~~~F_{\mu\nu}=\partial_\mu V_\nu -\partial_\nu V_\mu, \eqno{(2)} $$
where $m_v$, $g_s$ and $g_v$ are $\omega$-meson mass, $\sigma$-nucleon coupling
and $\omega$-nucleon coupling, respectively.
The potential $U(\phi )$ includes a nonlinear cubic-quartic terms of the scalar
field $\phi$; i.e.,
$$
U(\phi ) ={1\over{2}}m_s^2\phi^2+{{1}\over{3}}b\phi^3+{{1}\over{4}}c\phi^4,
\eqno{(3)} $$
where $m_s$ is $\sigma$-meson mass, and $b$ and $c$ are the constant parameters
which are determined phenomenologically.
The Lagrangian density (2) is also the same as the one used in the theory of
Boguta and Bodmer [13].

In the mean field theory of the point-like nucleon, at zero-temperature, the
baryon density $\rho_{pt}$ is given as
$$ \rho_{pt}={\lambda\over{3\pi^2}}k_F^3,    \eqno{(4)} $$
where $k_F$ is Fermi momentum and $\lambda =2$ in the nuclear matter.
In the model with the EVE [18][19][20], the volume $V$ for the $N$ body system
of nucleons in configurational space is reduced to the effective one, $V-NV_n$,
where $V_n$ is the volume of a nucleon.
According to this modification for the volume, the baryon density $\rho$ is
given by
$$ \rho={\rho'\over{1+V_n\rho'}}, \eqno{(5)} $$
where $\rho'$ has the same expression as $\rho_{pt}$ for the given $k_F$.
In a similar way, the scalar density is given by
$$ \rho_s={\rho'_s\over{1+V_n\rho'}} \eqno{(6)} $$
where $\rho'_s$ has the same expression as the scalar density of the system of
the point-like nucleons and is given by
$$
\rho_s'={\lambda\over{2\pi^2}}M^*[k_F\sqrt{k_F^2+M^{*2}}-M^{*2}\ln{({{k_F+\sqrt{k_F^2+M^{*2}}}\over{M^*}})}],    \eqno{(7)} $$
where $M^*$ is the effective nucleon mass.
{}From the equation of motion for the scalar meson, $M^*$ is given by
$$M^*=M-\Phi =M-{C_s^2\over{M^2}}(\rho_s-BM\Phi^2-C\Phi^3),   \eqno{(8)}  $$
where $C_s=g_sM/m_s$, and $\Phi /g_s$ is the  ground-state expectation value of
the field $\phi$.
The pressure $P$ and energy density $\epsilon$ are also given as
$$   P=P'+{{C_v^2}\over{2M^2}}\rho^2-U(\Phi ) \eqno{(9)} $$
and
$$   \epsilon
={{\epsilon'}\over{1+V_n\rho'}}+{{C_v^2}\over{2M^2 }}\rho^2
+U(\Phi )     \eqno{(10)} $$
, respectively, where $C_v=g_vM/m_v$,
$$
P'(k_F,M^*)={\lambda\over{12\pi^2}}\{E_F^*k_F(E_F^{*2}-{5\over{2}}M^{*2
})+{3\over{2}}M^{*4}\ln{({{E_F^*+k_F}\over{M^*}})}\},
\eqno{(11)} $$
and
$$   \epsilon'(k_F,M^*)=E_F^*\rho'-P'.   \eqno{(12)} $$
with $E_F^*=\sqrt{k_F^2+M^{*2}}$.
We remark that the pressure and the energy density of free point-like nucleons
can be given by eqs. (11) and (12), respectively, if we replace $M^*$ by the
free nucleon mass $M$.
The baryonic chemical potential is also given by
$$   \mu=E_F^*+V_nP'(k_F,M^*)+{{C_V^2}\over{M^2}}\rho.    \eqno{(13)} $$
It is easy to check that the equations (9),(10) and (13) satisfy the
thermodynamical identity [18], $\mu =(\epsilon +P)/\rho$, which yields the
following relation,
$$ M^*_0=\sqrt{E_{F0}^{*2}-k_{F0}^2}=[\{
M-a_1-V_nP'(k_{F0},M^*_0)-C_v^2\rho_0/M^2\} ^2-k_{F0}^2]^{1/2} \eqno{(14)} $$
at the normal density $\rho_0$, where $a_1$ is the binding energy of the normal
nuclear matter and the subscripts "0" denotes "at the normal density".
Since $P'(k_{F0},M^*_0)>0$, $\rho_0 >0$ and $C_v^2>0$, it is shown that
$$ M^*_0<\{ (M-a_1)^2-k_{F0}^2\}^{1/2}. \eqno{(15)} $$
This condition gives the upper bound for $M^*_0$. For example, if we put
$k_{F0}=1.4fm^{-1}$, we get $M^*_0<0.94M$.
(We remark that, due to the modification of baryon density (see eq. (5)),
$k_{F0}$ is larger in the theory with the EVE than in the theory of the
point-like nucleons, for the given $\rho_0$. )

The incompressibility $K$ at the normal density is defined as
$$ K=9\rho_0^2{{\partial^2e}\over{\partial
\rho^2}}\vert_{\rho=\rho_0}=9{{\partial P}\over{\partial
\rho}}\vert_{\rho=\rho_0}=9\rho_0{{\partial\mu}\over{\partial\rho}}\vert_{\rho=\rho_0},     \eqno{(16)} $$
where $e=\epsilon /\rho$.
In this paper, we also calculate the skewness coefficient, i.e., the
third-order derivative $K'$ of nuclear saturation curve.
In our definition, $K'$ is defined as
$$ K'=3\rho_0^3{{d^3 e}\over{d \rho^3}}\vert_{\rho=\rho_0}=3\rho_0{{d^2
P}\over{d \rho^2}}\vert_{\rho =\rho_0}-{{4}\over{3}}K
=3\rho_0^2{{d^2 \mu}\over{d\rho^2}}\vert_{\rho =\rho_0}-K. \eqno{(17)} $$
In this definition, large $K'$ means that the equations of state (EOS) is stiff
at high density.


There is the eight independent parameters in this model: i.e., $M$, $\rho_0$,
$a_1$, $R_n(=(3V_n/4\pi)^{1/3})$, $C_s$, $C_v$, $B$ and $C$.
In our calculations, we put $M=939$MeV, $\rho_0=0.17$fm$^{-3}$ and $a_1=16$MeV.
The other five parameters $R_n$, $C_s$, $C_v$, $B$ and $C$ are determined
phenomenologically.
Besides the two conditions for the saturation (i.e., $e_0=M-a_1$ and $P=0$), if
$M^*_0$, $K$ and $K'$ are given, we can determine the five parameters of the
model.
Conversely, we can calculate $M^*_0$, $K$ and $K'$, if the five parameters are
given.
As is seen in eq. (14), $M^*_0$ is determined if $R_n$ and $C_v$ are given.
Therefore, if we give one (two) quantity (quantities) in $R_n$, $C_v$ and
$M^*_0$ and give two (one) quantities (quantity) in $C_s$, $B$, $C$, $K$ and
$K'$, the other five quantities are automatically determined.
We also remark that we put $R_n=0\sim 0.9$fm in our calculations.
It is difficult to do the calculations at very large $R_n(> 0.9$fm), which is
close to $R_0(=\{3/(4\pi \rho_{0})\}^{1/3}\sim 1.1$fm), i.e., the half of the
averaged distance between two nucleons in the normal nuclear matter.

First we give $R_n$, $M^*_0$ and $K$, and calculate $K'$.
In fig. 1, we show $K'$ as a function of $R_n$ for the fixed values of $K$ and
$M^*_0$.
In the cases of small $K(\leq 200$MeV), $K'$ decreases in the large $R_n(\gsim
0.75$fm) region.
We remark that, in the case of $K=200MeV$ and $M^*_0=0.9M$, $K'$ does not
decrease much as in the other cases in fig.1(a).
$K'$ also decreases in the large $R_n$ region, in the cases of large $K(\geq
300$MeV) and small $M^*_0(=0.5M)$, although the absolute value of the decrease
is not large as in the cases of small $K$.
In the case of large $K$ and the large $M^*_0(=0.9M$), $K'$ hardly decreases.
The decrease of $K'$ is related to the fact that the coefficient $C$, which has
an important role for determining the high density behavior of EOS, becomes
(more) negative in the large $R_n$ region.
$C$ becomes (more) negative (i.e., attractive) to cancel the repulsive effect
of the EVE, to realize the fixed value of $K$, as $R_n$ increases.
( The decrease of $C$ is conspicuous in the cases of the small $M^*_0$, because
$C$ must also cancel the repulsive effects of the small $M^*_0$ as well as the
effects of the EVE.
Also in the cases of small $K$, the decrease of $C$ is conspicuous, because the
small $K$ must be reproduced. )
As a result, $K'$ becomes smaller in the large $R_n$ region, because of the
much negative $C$.
It is also seen that $K'$ always increases in the region of $R_n=0\sim 0.7$fm,
as $R_n$ increases, for the large $K(\geq 300$MeV).
In these cases, the absolute value of the increase is large for the large
$M^*_0(=0.9M)$.
Therefore, $K'$ is larger at $R_n=0.8$fm, which is often used as the nucleon
radius,
 than at $R_n=0$, in the cases of the large $K$ and the large $M^*_0$.
On the contrary, $K'$ is smaller at $R_n=0.8$fm than at $R_n=0$, in the cases
of the small $K$ and the small $M^*_0$.


\centerline{$\underline{~~~~~~~}$}

\centerline{Fig. 1(a),(b)}

\centerline{$\underline{~~~~~~~}$}


After $K'$ is determined, we can also calculate Coulomb coefficient $K_c$ in
the leptodermous expansion (1), using the scaling model, i.e., using the
following equation [1],
$$  K_c=-{{3q_{el}^2}\over{5R_0}}\biggl( {9K'\over{K}}+8 \biggr), \eqno{(18)}
$$
where $q_{el}$ is the electric charge of proton.
It is easily seen that $K_c$ becomes more negative as $K'$ increases in eq.
(18), if $K$ is fixed.
$K_c$ is negative and the absolute value is large in the EOS, which becomes
stiffer at high density.
In fig. 2, we show $K_c$ as a function of $K$ with the fixed values of $R_n$
and $M^*_0$.
In the case of $M^*_0/M=0.5(0.7, 0.9)$, $K_c$ is smaller (more negative) at
$R_n=0.8$fm than at $R_n=0$, if $K\gsim 250(260, 160)$MeV.
The reason is  that, in these cases, $K'$ is larger at $R_n=0.8$fm than at
$R_n=0$, as is seen in fig. 1.
It is remarkable that, in the case of $M^*_0/M=0.5(0.7, 0.9)$, the $K_c$ is
larger (less negative) at $R_n=0.8$fm than at $R_n=0$, if $K\lsim 250(260,
160)$MeV.
Naturally, the reason is that, in these cases, $K'$ is smaller at $R_n=0.8$fm
than at $R_n=0$, as is also seen in fig. 1.
The EVE make $K'$ smaller and make $K_c$ less negative for the small $K$.
This change of $K_c$ makes it possible to fit the empirical values of $K_c$ for
the small $K$.
We also remark that the change of $K_c$ by the EVE is opposite to the one by
the vector meson self-interaction (VSI) [12].
It is reasonable that the repulsive EVE has opposite effects to the ones by the
attractive VSI.


\centerline{$\underline{~~~~~~~}$}

\centerline{Fig. 2}

\centerline{$\underline{~~~~~~~}$}


If we give $K$, $K_c$ and $R_n$, the other parameters of the model are also
determined.
( We remark that $K'$ is also uniquely determined by eq. (18), if $K$ and $K_c$
are given. )
In table 2, we show the examples of the parameters sets which reproduce the
empirical values of $K$ and $K_c$ in table 1 with several values of $R_n$.
For simplicity, we use the average values of $K$ and $K_c$ in table 1.


\centerline{$\underline{~~~~~~~}$}

\centerline{Table 2(a),(b),(c),(d)}

\centerline{$\underline{~~~~~~~}$}


We could not find the parameter sets which reproduce $(K,
K_c)=(200.0,2.577)$MeV at any $R_n$, as in the case of $R_n=0$ [11], where the
calculated $K_c$ is always smaller than the empirical value $2.577$MeV.
This fact is understood as follows.
At $R_n=0$ and $K=200$MeV, the largest $M^*_0$ has the largest $K_c$ which is
closest to the empirical value.
However, as is seen in fig. 1(a), $K'$ does not become much smaller in the
cases of $K=200$MeV and large $M^*_0 (\gsim 0.9M$), even if we use a very large
$R_n(>0.8)$fm.
As a result, in those cases, $K_c$ hardly becomes large and does not reproduce
the empirical value.
On the other hand, in the cases of $K=200$MeV and $M^*_0~\lsim 0.7M$, the value
of $K_c$ is much smaller at $R_n=0$ than the empirical value.
In those cases, although the EVE makes $K'$ much smaller and makes $K_c$ much
larger in the large $R_n$ region, $K_c$ is still far from 2.577MeV.

In the case of $(K,K_c)=(250.0, -0.7065)$MeV, the empirical values is
reproduced by this model in the regions of $R_n\approx 0\sim0.6$fm and
$R_n=0.85 \sim 0.88$fm.
The calculated effective nucleon mass $M^*_0$ increases as $R_n$ increases in
the former region.
The solution of parameter set disappears at $R_n=0.6$fm, because of the upper
bound condition (15) for $M^*_0$.
The solution appears again at $R_n=0.85$fm.
This reappearance of the solutions is related to the fact that $K'$ decreases
in the large $R_n$ region.
The disappearance of the solution in the intermediate region of $R_n$ occurs
also in the cases of $(K,K_c)=$(300.0, -3.990)MeV and (350.0,-7.274)MeV.
The solution disappears at $R_n=0.72(0.79)$fm and reappears at
$R_n=0.86(0.85)$fm in the case of $(K,K_c)=$(300.0, -3.990)MeV
((350.0,-7.274)MeV).

In the cases of $(K,K_c)$=(150.0,5.861)MeV and (143,3.04)MeV,
at $R_n=0$, there is no parameter set, which reproduce the empirical values.
However, we could find the parameter sets for $(K,K_c)=$(150.0,5.861)MeV and
for$(K,K_c)$=(143, 3.04)MeV, if we put $R_n=0.81\sim0.84$fm and $R_n=0.76\sim
0.87$fm, respectively.
(See EOS10 and EOS11 in table 2(d). )
The reason is that, as is seen in fig. 1, the EVE makes $K'$ much smaller in
the large $R_n(\lsim 0.75$fm) region, in the case of $K\sim 150$MeV.

We remark that the coefficient $C$ is always negative in the these solutions,
as is seen in the EOS 10 and EOS 11.
$C$ becomes (more) negative (i.e., attractive) to cancel the large repulsive
effects of the large $R_n$, as is mentioned before.
Negative $C$ may cause the difficulty such as a bifurcation of solution of
$\Phi$.
However, this difficulty is modified by introducing the higher terms of $\Phi$
which hardly affect the nuclear matter properties at the normal density.
For example, we add the following terms to $U(\Phi )$.
$$    U_{56}(\Phi )={D\over{5M}}\Phi^5+{E\over{6M^2}}\Phi^6, \eqno{(19)} $$
where $D$ and $E$ are the dimensionless constants.
If we put $D=-0.00755$ and $E=0.006$, for example, we get the EOS 12 in table
2(d).
Although, at $\rho =\rho_0$, the EOS 12 has almost the same properties as in
the EOS 11, it has only one solution as is seen fig 3, where $dU(\Phi )/d\Phi
-\rho_s$ is shown as a function of $\Phi$, both in the cases of the EOS 11 and
of the EOS 12. (We remark $dU(\Phi )/d\Phi -\rho_s=0$ is a equivalent condition
to eq. (8). )

We also calculate the volume-symmetry coefficient $K_{vs}$ in the expansion
(1).
Because the $\rho$-meson effects are important in the symmetry properties
[21][22], we add the standard $\rho$-meson terms to the Lagrangian
[21][22][19].
According to the modification, the following term is added to the energy
density [21][22][19].
$$ \epsilon_\rho
={{g_\rho^2}\over{8m_\rho^2}}(\rho_p-\rho_n)^2={{C_\rho^2}\over{8M^2}}\rho_3^2,~~~~~~~~\rho_3=\rho_p-\rho_n, \eqno{(20)} $$
where $m_\rho$, $g_\rho$, $\rho_p$ and $\rho_n$ are $\rho$ meson mass,
$\rho$-nucleon coupling, proton density and neutron density, respectively, and
$C_\rho =g_\rho M/m_\rho$.
In the theory with the EVE [19], $\rho_p$ and $\rho_n$ are given by
$$ \rho_p={\rho_p'\over{1+V_n\rho'}}~~~~~{\rm
and}~~~~~\rho_n={\rho'_n\over{1+V_n\rho'}}, \eqno{(21)} $$
, where the expressions for $\rho'_p$ and $\rho'_n$ have the same expressions
as the densities of the point-like proton and neutron, respectively.
In the mean field theory, the inclusion of the $\rho$-meson effects do not
affect the saturation conditions and do not change the properties such as $K$
and $K_c$ in the symmetric nuclear matter.
Therefore, the determination of the parameters $R_n$, $C_s$, $C_v$, $B$ and $C$
in fitting the data of $K$ and $K_c$ is not affected by this modification of
the Lagrangian.
Using the modified Lagrangian with the parameter sets in table 2, we calculate
$K_{vs}$ with aid of the scaling model [1]: i.e.,
$$  K_{vs}=K_{sym}-L\biggl( 9{K'\over{K}}+6 \biggr), \eqno{(22)} $$
where
$$  L=3\rho_{0}{{d a_{4}}\over{d \rho}}\vert_{\rho =\rho_{0}},
{}~~~~K_{sym}=9\rho_{0}^2{{d^2 a_{4}}\over{d \rho^2}}\vert_{\rho =\rho_{0}},
{}~~~~{\rm and}
{}~~~~a_{4}={1\over{2}}\rho{{\partial^2\epsilon}\over{\partial\rho_3^2}}\vert_{\rho_3=0}.
  \eqno{(23)}  $$
The results are also summarized in table 2.
In these calculations, we determine the $\rho$ meson coupling $g_\rho$ so as to
realize $a_4=30.0$MeV at $\rho =\rho_0$.
By comparing the table 1 and table 2, it is seen that $K=300\pm50$MeV is
favorable to account for the empirical values of $K$, $K_c$ and $K_{vs}$
simultaneously, as in the case of $R_n=0$ [11], and as in the case of the
mean-field theory with the VSI [12].
This unchanged conclusion is related to the fact that $K_{vs}$ is very
sensitive to the ratio $K'/K$, which is adjusted to the empirical value in
table 2.
It seems that this feature is not changed drastically, if we use the
relativistic mean-field theory and the scaling model.

In the last, we examine whether the QMC result is well reproduced by the EVE
model.
According to ref. [17], $K=200$MeV and $M^*_0=0.906M$, if the bag radius
$R_B=0.8$fm is used.
Since the value of $K'$ and $K_c$ is not shown in the reference, we calculate
$K'$ by using the fig. 2 in ref. [17], where the saturation curve of the
nuclear matter in the QMC model is shown.
The result is $K'\sim  -85$MeV.
(This value corresponds to $K_c\sim -3.2$MeV, which is somewhat larger than the
empirical value, $2.577\pm 2.06$MeV. )
We search the parameter set of the EVE model, which reproduces the $K=200$MeV,
$M^*_0=0.906M$ and $K'=-85$MeV.
The results are shown in table 3.
In the parameter set, $R_n(=0.568$fm) is somewhat smaller than $R_B(=0.8$fm).
The physical meaning of $R_n$ in the EVE model may be somewhat different from
that of the bag radius $R_B$ in the QMC model.


\centerline{$\underline{~~~~~~~}$}

\centerline{Table 3}

\centerline{$\underline{~~~~~~~}$}



In summary, we have studied the compressional properties of nuclear matter by
using the relativistic mean field theory with the nonlinear $\sigma$ terms and
the EVE of the nucleons, under the assumption of the scaling model.
We found that the EVE yields the possibility to reproduce the empirical values
of $K_c$ with the small $K\sim 150$MeV.
However, if we require that $K_{vs}$ should be reproduced as well as $K_c$, $K=
300\pm 50$MeV is favorable.
It seem to be difficult to change this conclusion drastically, in the framework
of the relativistic mean field theory and the scaling model.

\noindent
$Acknowledgment$: The authors are grateful to N. Kakuta for useful discussions,
and to the members of nuclear theorist group in Kyushyu district in Japan for
their continuous encouragement.
One (H. K.) of the authors would like to thank K. Saito for informing his
publications.
The authors also gratefully acknowledge the computing time granted by the
Research Center for Nuclear Physics (RCNP).

\vfill\eject

\centerline{{\bf{References}}}

\noindent
[1] J.P. Blaizot, Phys. Rep. {\bf{64}}(1980)171.

\noindent
[2] J.M. Pearson, Phys. Lett. B{\bf{271}}(1991)12.

\noindent
[3] S. Shlomo and D.H. Youngblood, Phys. Rev. {\bf{C47}}(1993)529.

\noindent
[4] S. Rudaz, P.J. Ellis, E.K. Heide and M. Prakash, Phys. Lett.
{\bf{B285}}(1992)183.

\noindent
[5] E.K. Heide and S. Rudaz, Phys. Lett. {\bf{B262}}(1991)375.

\noindent
[6] D. Von-Eiff, J.M. Pearson, W. Stocker and M.K. Weigel,

\noindent
Phys Lett. {\bf{B324}}(1994)279.

\noindent
[7] D. Von-Eiff, J.M. Pearson, W. Stocker and M.K. Weigel,

\noindent
Phys. Rev. {\bf{C50}}(1994)831.

\noindent
[8] D. Von-Eiff, W. Stocker and M.K. Weigel, Phys. Rev. {\bf{C50}}(1994)1436.

\noindent
[9] A.R. Bodmer and C.E. Price, Nucl. Phys. {\bf{A505}}(1989)123.

\noindent
[10] M.V. Stoitsov, P. Ring and M.M. Sharma, Phys. Rev. {\bf{C50}}(1994)1445.

\noindent
[11] H. Kouno et al., Phys. Rev. {\bf{C51}}(1995)1754.

\noindent
[12] H. Kouno et al., to be published in Phys. Rev. C.

\noindent
[13] J. Boguta and A.R. Bodmer, Nucl. Phys. {\bf{A292}}(1977)413.

\noindent
[14] A.R. Bodmer, Nucl. Phys. {\bf{A526}}(1991)703.

\noindent
[15] P.A.M. Guichon, Phys. Lett. {\bf{B200}}(1988)235.

\noindent
[16] K. Saito and A. W. Thomas, Nucl. Phys. , {\bf{A574}}(1994)659.

\noindent
[17] K. Saito and A. W. Thomas, Phys. Lett., {\bf{B327}}(1994)9.

\noindent
[18] D.H. Rischke, M.I. Gorenstein, H. St{\"{o}}cker and W. Greiner,

\noindent
Z. Phys., {\bf{C51}}(1991)485.

\noindent
[19] Qi-Ren Zhang, Bo-Qiang Ma and W. Greiner, J. Phys. {\bf{G18}}(1992)2051.

\noindent
[20] Bo-Qiang Ma, Qi-Ren Zhang, D.H. Rischke and W. Greiner,

\noindent
Phys. Lett., {\bf{B315}}(1993)29.

\noindent
[21] B.D. Serot, Phys. Lett. {\bf{86B}}(1979)146

\noindent
[22] B.D. Serot and J.D. Walecka, $The$ $Relativistic$ $Nuclear$ $Many$-$Body$
$Problem$ in: Advances in nuclear physics, vol. 16 (Plenum Press, New York,
1986).


\vfill\eject

\centerline{{\bf{Table and Figure Captions}}}

\bigskip

\noindent
Table 1

\noindent
The sets of the empirical values of $K$, $K_c$ and $K_{vs}$.
(Shown in MeV.)
The sets $1\sim 5$  is the data from the table 3 in ref. [2]. (According to the
conclusion in ref. [2], we only show the data in the cases of $K=150\sim
350$MeV.) The set 6 is one example of the data of the table IV in ref. [3].

\bigskip

\noindent
Table 2

\noindent
Parameter sets fitted for the mean values of the empirical values of $K$ and
$K_c$ in table 1.
The $K$, $K_c$, $K'$, $K_{vs}$, $L$ and $K_{sym}$ are shown in MeV, while $R_n$
are shown in fm.
In EOS12, D=-0.00755 and E=0.006. (See eq. (19).)

\bigskip

\noindent
Table 3

\noindent
Parameter sets fitted for $K=200$MeV, $M^*_0=0.906M$, and $K'=-85$MeV.
$R_n$ are shown in fm.

\bigskip

\noindent
Fig. 1~~~~~$K'$ as a function of $R_n$ with several values of $M^*_0$ and $K$.
In the figure (a) ((b)), the solid line, the dotted line, and the dash-dotted
line are the results with $M^*_0/M=$0.5, 0.7, and 0.9, respectively, for
$K=150(300)$MeV.
In the figure (a) ((b)), the bold solid line, the bold dotted line, and the
bold dash-dotted line are the results with $M^*_0/M=$0.5, 0.7, and 0.9,
respectively, for $K=200(400)$MeV.

\bigskip

\noindent
Fig. 2~~~~~$K_c$ as a function of $K$ with fixed values of $R_n$ and $M^*_0$.
The solid line, the dotted line, and the dash-dotted line are the results with
$M^*_0/M=$0.5, 0.7, and 0.9, respectively, in the case of $R_n=0$.
The bold solid line, the bold dotted line, and the bold dash-dotted line are
the results with $M^*_0/M=$0.5, 0.7, and 0.9, respectively, in the case of
$R_n=0.8$fm.
The crosses with error bars are the data sets $1\sim 5$ in table 1.
The solid squares are the data from the table IV in ref. [3].
(For simplicity of the figure, we omit the error bars in the latter data. )

\bigskip

\noindent
Fig. 3~~~~~$dU(\Phi )/d\Phi-\rho_s$ as a function of $\Phi$.
The solid line and the dotted line are the results of EOS11 and EOS12,
respectively.

\vfill\eject


\large

\hspace*{-3.5cm}
  \begin{tabular}{ccccccc}
                                    \hline
    \     & Set 1 & Set 2 & Set 3 & Set 4 & Set 5  & Set 6 \\ \hline
    \   $K$ & 150.0 & 200.0  & 250.0 & 300.0 & 350.0 &$143\pm53$ \\
    \  $K_c$ & $5.861\pm2.06 $ & $2.577\pm2.06 $ & $-0.7065\pm2.06 $
     & $-3.990\pm2.06 $ &$-7.274\pm2.06 $ & $3.04\pm4$ \\
    \ $K_{vs}$ & $66.83\pm101$ & $-46.94\pm101$ & $-160.7\pm101$
     & $-274.5\pm101$ & $-388.3\pm101$ & $34\pm159$ \\ \hline
  \end{tabular}

\bigskip

\begin{center}
Table 1
\end{center}

\bigskip


\large

\begin{center}
  \begin{tabular}{cccccccc}
                                    \hline
    \  $R_n$ & $C_s^2$ & $C_v^2$ & $B$ & $C$ \\ \hline
    \  0.568 & 106.17 & 16.472 & -2.589$\times 10^{-2}$ & 1.049 \\ \hline
  \end{tabular}

\bigskip

Table 3
\end{center}

\bigskip

\vfill\eject


\large

  \begin{tabular}{cccc}
                                    \hline
    \    EOS    & 1      & 2    & 3   \\ \hline
    \ $K $ & 250.0  &  300.0    & 350.0 \\
    \ $K_c $ & -0.7065 & -3.990 & -7.274 \\
    \ $K^\prime $ & -197 & -94.0 & 56.0 \\
    \ $R_n$      & 0.00 & 0.00 & 0.00  \\
    \ $ M_0^\ast/ M $ & 0.910 & 0.831 & 0.609 \\
    \ $K_{vs}$ & 53.83 & -275.5 & -626.8 \\
    \ $L$ & 76.61 & 78.90 & 93.81  \\
    \ $K_{ sym}$ &-29.20 & -24.68 & 71.27 \\
    \ $ C_s^2 $ & 42.392 & 144.56 & 293.29 \\
    \ $ C_v^2 $ & 18.459 & 66.337 & 197.85 \\
    \ $ B $ & -0.5282 & -5.067$\times 10^{-3}$ & 1.595$\times 10^{-3}$ \\
    \ $ C $ & ~~~5.071~~~ & ~~~0.1028~~~ &-1.686$\times 10^{-3}$ \\
    \ $ C_\rho^2 $ & 90.31 & 84.11 & 59.94 \\ \hline
  \end{tabular}

\bigskip

\begin{center}
      Table 2(a)
\end{center}



\large

  \begin{tabular}{cccc}
                                    \hline
    \    EOS    & 4      & 5    & 6   \\ \hline
    \ $K $ & 250.0  &  300.0    & 350.0 \\
    \ $K_c $ & -0.7065 & -3.990 & -7.274 \\
    \ $K^\prime $ & -197 & -94.0 & 56.0 \\
    \ $R_n$      & 0.60 & 0.72 & 0.79  \\
    \ $ M_0^\ast/ M $ & 0.931 & 0.919 & 0.902 \\
    \ $K_{vs}$ & 77.41 & -244.9 & -600.1 \\
    \ $L$ & 79.75 & 84.81 & 91.54  \\
    \ $K_{ sym}$ & -9.193 & 24.76 & 80.95 \\
    \ $ C_s^2 $ & 18.121 & 55.348 & 100.08 \\
    \ $ C_v^2 $ & 0.42953 & 2.0916 & 6.7682 \\
    \ $ B $ & -2.063 & -0.2924 & 3.295$\times 10^{-2}$ \\
    \ $ C $ & ~~~23.00~~~ & ~~~3.728~~~ & ~~~0.2168~~~ \\
    \ $ C_\rho^2 $ & 83.83 & 75.39 & 66.67 \\ \hline
  \end{tabular}

\bigskip

\begin{center}
      Table 2(b)
\end{center}

\vfill\eject


\large

  \begin{tabular}{cccc}
                                    \hline
    \    EOS    & 7      & 8    & 9   \\ \hline
    \ $K $ & 250.0  &  300.0    & 350.0 \\
    \ $K_c $ & -0.7065 & -3.990 & -7.274 \\
    \ $K^\prime $ & -197 & -94.0 & 56.0 \\
    \ $R_n$      & 0.85 & 0.86 & 0.85  \\
    \ $ M_0^\ast/ M $ & 0.573 & 0.540 & 0.643 \\
    \ $K_{vs}$ & 1054 & 597.5 & -407.3 \\
    \ $L$ & 150.5 & 164.1 & 133.1  \\
    \ $K_{ sym}$ & 890.6 & 1119 & 582.8 \\
    \ $ C_s^2 $ & 289.99 & 302.02 & 245.09 \\
    \ $ C_v^2 $ & 181.40 & 195.55 & 144.14 \\
    \ $ B $ & 2.785$\times 10^{-3}$ & 2.213$\times 10^{-3}$ & 3.421$\times
10^{-3}$ \\
    \ $ C $ & -7.121$\times 10^{-3}$ & -5.987$\times 10^{-3}$ &-9.667$\times
10^{-3}$ \\
    \ $ C_\rho^2 $ & 11.18 & 2.037 & 23.87 \\ \hline
  \end{tabular}

\bigskip

\begin{center}
      Table 2(c)
\end{center}



\large

  \begin{tabular}{cccc}
                                    \hline
    \    EOS    & 10      & 11    & 12   \\ \hline
    \ $K $ & 150.0  &  143.0   & 143.0 \\
    \ $K_c $ & 5.861 & 3.04 & 3.04 \\
    \ $K^\prime $ & -260 & -190 & -190 \\
    \ $R_n$      & 0.81 & 0.76 & 0.76  \\
    \ $ M_0^\ast/ M $ & 0.574 & 0.623 & 0.623 \\
    \ $K_{vs}$ & 1988 & 1060 & 1060 \\
    \ $L$ & 137.4 & 116.9 & 116.9 \\
    \ $K_{ sym}$ & 668.8 & 364.5 & 364.5 \\
    \ $ C_s^2 $ & 307.94 & 291.23 & 287.83 \\
    \ $ C_v^2 $ & 189.47 & 170.39 & 170.39 \\
    \ $ B $ & 3.370$\times 10^{-3}$ & 4.527$\times 10^{-3}$ & 3.881$\times
10^{-3}$ \\
    \ $ C $ & -7.261$\times 10^{-3}$ & -8.976$\times 10^{-3}$ &-5.556$\times
10^{-3}$ \\
    \ $ C_\rho^2 $ & 19.83 & 36.22 & 36.22 \\ \hline
  \end{tabular}

\bigskip

\begin{center}
      Table 2(d)
\end{center}


\end{document}